# Theoretical kinetic studies of Venus chemistry. Formation and destruction of SCl, $SCl_2$, and HSCl

David E. Woon*

Department of Chemistry, University of Illinois at Urbana-Champaign, 600 S. Mathews Avenue, Urbana, IL 61801

Dominique M. Maffucci and Eric Herbst

Department of Chemistry, University of Virginia, P.O. Box 400319, Charlottesville, VA 22904

Department of Astronomy, University of Virginia, P.O. Box 400325, Charlottesville, VA 22904

## ABSTRACT

Accurate and thorough characterization of the chemistry of compounds containing the third-row elements sulfur and chlorine is critical for modeling the composition of the atmosphere of Venus. We have used a combination of ab initio quantum chemistry and kinetic theory to characterize a group of nine exothermic reactions that involve the exotic sulfur-chlorine species SCl, $SCl_2$, and HSCl, which are thought to be present in trace quantities in the atmosphere of Venus and are included to various degrees in the published atmospheric models. Reaction pathways were characterized with coupled cluster theory at the RCCSD(T) level with triple $\zeta$ quality correlation consistent basis sets. For reactions with barriers that lie above the reactant asymptote, the barrier height was extrapolated to the RCCSD(T) complete basis set level via single-point calculations with quadruple and quintuple ζ quality sets. Rate coefficients were predicted with capture theory and transition state theory as appropriate. We have found that in some cases addition-elimination reactions can compete with abstraction reactions due to the tendency of sulfur to form hypervalent compounds and intermediates via recoupled pair bonding.

## 1. Introduction

Although compounds containing sulfur are only present in trace amounts in the Venusian atmosphere, which is dominated by carbon dioxide and molecular nitrogen, they nevertheless profoundly impact the character and aspect of the planet. Most notably, the surface of Venus is obscured by a dense canopy of sulfuric acid aerosol clouds (Bézard and de Bergh, 2007; Gao et al., 2013). In addition to sulfuric acid ($H_2SO_4$) (Hansen & Hovenier, 1974), other sulfur compounds known to be present in the atmosphere of Venus include sulfur dioxide ($SO_2$) (Barker, 1979; Stewart et al., 1979), sulfur monoxide (SO) (Na et al., 1990), carbonyl sulfide (OCS) (Bézard et al., 1990), a tentative detection of $S_3$ (Maiorov et al., 2005), and the recent detection of two rotamers of OSSO (Frandzen, Wenneberg, and Kjaergaard, 2016). Although the detection of hydrogen sulfide ($H_2S$) was reported on the basis of data from both the Pioneer Venus probe (Hoffman et al. 1980) and the Venera 13 and 14 landers (Mukhin et al. 1983, cited by Krasnopolsky, 2008), a ground-based, high-resolution study placed an upper limit on $H_2S$ of just 23 ppb (Krasnopolsky, 2008).

The current models that aim to account for the composition of Venus' atmosphere (Krasnopolsky, 2006, 2006, 2012; Krasnopolsky and Lefèvre, 2013; Marcq et al., 2018; Mills and Allen, 2007; Mills et al., 2007; Zhang et al., 2012) include a number of other sulfur compounds that are plausibly thought to be present, including $SO_3$, S, $S_2$, $S_4$–$S_8$, $H_2S$, SH, $HSO_3$, $S_2O_2$, and $S_2O$. Also included in the models are a small number of compounds that contain both sulfur and chlorine, including SCl, $SCl_2$, HSCl, OSCl, $ClSO_2$, and $SO_2Cl_2$. To date, the only chlorine compounds detected on Venus are HCl (Connes et al., 1967) and ClO (Sandor and Clancy, 2013).

As part of an extended study of Venus chemistry, we have characterized a set of nine formation and destruction reactions involving the sulfur- and chlorine- containing species SCl, $SCl_2$, and HSCl. The primary objective of our project is to use theoretical chemical methods to predict accurate reaction rate coefficients for reactions involving sulfur-containing compounds that are thought to be plausibly present in the atmosphere of Venus. These reactions are often represented in the models with approximate rate coefficients due to the absence of measured values. We also seek to identify additional sulfur-containing compounds (and their reactions) that are not presently included in the published models.

Most of the bimolecular reactions presently included in Venus modeling studies are either abstraction reactions of the general form

$$AB + C \rightarrow A + BC \tag{1}$$

or addition-elimination reactions of the form

$$AB + C \rightarrow ABC \rightarrow A + BC \tag{2}$$

where ABC is a stable intermediate that tends to decompose due to the excessive internal energy with which it is formed. An intermediate may also isomerize into another intermediate before decomposing. In some circumstances, intermediates can be stabilized. This can occur (i) if the pressure is large enough that it's efficient for bath gas molecules to carry away sufficient excess energy to prevent elimination or (ii) if the intermediate is large enough to relax vibrationally before elimination can occur. Abstraction reactions involve no intermediates. As written, the two reactions above have the same products, but both the energetics and the kinetics of the two reactions can be quite different. Abstraction reactions frequently have reaction barriers that impede the progress of the reactions and result in slow reaction rates. Addition-elimination reactions often have no barriers in the same system and thus can be more efficient than abstraction reactions.

In two cases, we have identified new pathways where addition-elimination reactions compete with abstraction reactions, where neither pathway has been included in any of the published models. Sulfur is widely known to undergo hypervalent bonding, which occurs because it is able to form recoupled pair bonds (Woon and Dunning, 2009; Dunning et al., 2013). Consideration of hypervalency leads to the identification of new addition-elimination pathways that proceed through a stable hypervalent intermediate, SABC:

$$\mathrm{SAB + C \rightarrow SABC \rightarrow SAC + B} \quad (3)$$

While the S-containing reactant and product are typical divalent S compounds in this reaction, the SABC intermediate is hypervalent. We will treat two such cases in this work involving hypervalent intermediates: the H + HSCl and OH + HSCl reactions.

In the remainder of this paper, we will first describe the methodologies used to generate reaction pathways and the associated rate coefficients for pressures and temperatures relevant to the upper, middle, and lower atmosphere of Venus (100–800 K). We will then describe our results for nine exothermic reactions and will briefly discuss their endothermic reverse reactions. In several cases, there are measured values of the rate coefficients with which we can compare our results though usually the rate coefficients were measured at conditions outside of what is present on Venus. A brief section of conclusions will end the paper.

## 2. Methods

### *2.1. Quantum chemical calculations*

In this work we used high level coupled cluster theory to characterize minima and transition states on reaction pathways using the MOLPRO code (Werner et al., 2008). The accuracy of correlated quantum chemical methods depends upon how completely they treat excitations from the reference function (usually the single-reference Hartree-Fock wavefunction or a multireference wavefunction built by adding a limited number of low-lying excitations to the Hartree-Fock configuration). In this work we used the level of coupled cluster theory that correlates the Hartree-Fock wavefunction with full treatment of single and double excitations and an approximate, perturbative treatment of triple excitations that is designated RCCSD(T) (Purvis and Bartlett, 1982; Raghavachari et al., 1989; Knowles, Hampel, & Werner, 1993; Watts, Gauss, & Bartlett, 1993). RCCSD(T) theory is considered to be the "gold standard" for quantum chemistry, because it balances accuracy and tractability (Lee and Scuseria, 1995; Bartlett and Musiał, 2007; Řeźač and Hobza, 2013). The occupied and virtual molecular orbitals used in the RCCSD(T) calculations are expanded in a one-electron basis set. Here, we mostly used correlation consistent basis sets at the triple zeta (3ζ) level that include diffuse basis functions that are useful for describing long-range interactions and other properties of molecules as well as an extra tight d function on S and Cl atoms (Dunning, 1989; Kendall, Dunning, and Harrison, 1992; Woon and Dunning, 1993; Dunning, Peterson, and Wilson, 2001). The full designations for these basis sets are aug-cc-pVTZ for H and O and aug-cc-pV(T+d)Z for S and Cl. We will use the shorthand notation "AVTZ" henceforth to designate this level of basis set. These sets include functions as high as f angular momentum for the three heavy atoms. We also performed limited additional calculations at the quadruple zeta (4ζ) and quintuple zeta (5ζ) levels, for which we will use the shorthand notations "AVQZ" and "AV5Z" (QZ sets add additional spdf primitives as well as g functions; 5Z is the first level that includes h functions). These consisted of single point energy calculations at the RCCSD(T)/AVQZ//RCCSD(T)/AVTZ and RCCSD(T)/AV5Z//RCCSD(T)/AVTZ levels (where X//Y denotes a calculation at level X using the optimized geometry of a calculation at level Y). We will simplify the notation somewhat and refer to these as RCCSD(T)/AVQZ-sp and RCCSD(T)/AV5Z-sp calculations, where the RCCSD(T)/AVTZ geometry is implied. In this study, only valence electrons are correlated.

Harmonic frequencies were calculated for stationary points (minima and transition states) at the RCCSD(T)/AVTZ level by expanding a potential energy surface (PES) around a putative structure by displacements of the 3(*natoms*)–6 internal coordinates, including couplings. The expansion includes all of the linear, quadratic, and cubic terms plus even quartic terms. Frequency calculations serve to verify that a geometric structure is a true stationary point with all the first derivatives acceptably close to zero and either a minimum (second derivatives that are all positive) or transition state (TS, second derivatives that are all positive except for one vibrational mode). The frequencies, including the imaginary one for a transition state, were used in the rate calculations as well as to correct the equilibrium energies of the stationary points for their zero-point energies (ZPE). For our calculations where H is replaced with D, we reran the calculations of the frequencies with the higher mass of D, which shifts frequencies associated with D motion downward and thus reduces the net ZPE for the species.

For reactions with positive barriers, single point calculations were performed with AVQZ and AV5Z basis sets so that the barrier heights could be extrapolated to the estimated complete basis set (CBS) limit for the RCCSD(T) level. The extrapolations were performed on the total energies rather on the barrier heights. The form of equation used for extrapolation is a simple exponential function, $f(x) = A_{CBS} + B\mathrm{e}^{Cx}$, where $x$ is the basis set ordinal and $A_{CBS}$ is the estimated complete basis set limit (Woon, 1993). These are identified as RCCSD(T)/CBS-sp level calculations, since the structures were not reoptimized beyond the AVTZ level.

The rate calculations also required other properties that were computed quantum chemically. Bimolecular collision rates depend upon the permanent electrostatic moments and the various polarizabilities of molecules. Values for the dipole moments and dipole polarizabilities were calculated at the RCCSD(T)/AVTZ level for all of the reactant molecules using the finite field approach, which we have described in detail elsewhere (Woon and Herbst, 2009; Müller and Woon, 2013). Also needed were the ionization energies (IE) of the reactants, which were taken from published experimental data (NIST

WebBook; Linstrom & Mallard, 2018) when available. For the small number of compounds with no measured value, we calculated the IE at the RCCSD(T)/CBS-sp level. For SH (where there's a measured value), this yields only about 0.05 eV error. The structures of the associated cations were optimized at the RCCSD(T) level. ZPE corrections were included.

The level of quantum chemical theory used here is quite robust (especially for the barrier heights). Coupled cluster theory has been extended beyond the RCCSD(T) level to include full treatment of triple and quadruple excitations, RCCSDTQ, but it is very intensive computationally. It is likely, however, that this is the largest source of error remaining in the calculations. Other approximations are expected to have less impact, such as correlating only the valence electrons instead of all of the electrons.

*2.2. Reaction rate theory*

*2.2.1. Exothermic reactions with a single barrier*

For reactants A and B and transition state ‡, the transition state theory reaction rate coefficient (Woon and Herbst, 1996) is

$$k^{\mathrm{TST}}(T) = \frac{k_B T}{h} \frac{Q^{\ddagger}}{Q_A Q_B} e^{-E_0/k_B T} \tag{4}$$

where $E_0$ is the difference between the reactant system and the transition state zero-point energies, $k_B$ is the Boltzmann constant, $h$ is the Planck constant, and $c$ is the speed of light. The molecular partition functions

$$Q = Q_{\mathrm{rot}} Q_{\mathrm{vib}} Q_{\mathrm{trans}} Q_{\mathrm{elec}} \tag{5}$$

for the reactants and transition state are the products of contributions from the rotational, vibrational, translational, and electronic degrees of freedom.

We parameterize the partition functions by equivalent temperatures to observe by inspection the relative contributions of each rotational and vibrational mode to the molecular partition function over the temperature range $T$ = 10–800 K, which includes the range of temperatures used in chemical models of the Venusian atmosphere. In terms of the rotational constants $\tilde{B} = B/hc$, where $B = h/8\pi^2 cI$ and $I$ is the moment of inertia, the rotational temperature for a linear molecule becomes $\theta_r = hc\tilde{B}/k_B$. When energy levels are closely spaced, the partition function for linear molecules can be integrated to yield

$$Q_{\mathrm{rot}} = \frac{g_{\mathrm{I}}^{\mathrm{ns}}}{\sigma} \sum_J (2J+1) e^{-BJ(J+1)/k_B T} \approx \frac{g_{\mathrm{I}}^{\mathrm{ns}}}{\sigma} \frac{k_B T}{B} = \frac{\boldsymbol{g_{\mathrm{I}}^{\mathrm{ns}}}}{\boldsymbol{\sigma}} \frac{\boldsymbol{k_B T}}{\boldsymbol{hc\tilde{B}}} = \frac{g_{\mathrm{I}}^{\mathrm{ns}}}{\sigma} \frac{T}{\theta_r} \tag{6}$$

and written in terms of the rotational temperature $\theta_r$. The symmetry number $\sigma$ accounts for indistinguishable configurations of identical nuclei (see Fernández-Ramos et al., 2007) and is equal to 2 for homonuclear diatomic species such as $Cl_2$, and 1 for heteronuclear diatomic species such as SCl and OH. In general, the symmetry number $\sigma$ for a molecule in a given point group is equal to the ratio of the number of permutations of equivalent nuclei to the number of configurations unique under rotation. The rotational symmetry places constraints on the allowed symmetries of the total molecular wavefunction, so the nuclear degeneracies and symmetries must be considered and appropriately coupled to rotational states such that the total molecular nuclear spin wavefunction exhibits appropriate behavior under the parity operator. For heteronuclear diatomic species, the nuclear spin degeneracy factor $g_I^{\mathrm{ns}} = (2I_a + 1)(2\boldsymbol{I_b} + 1)$ accounts for all possible nuclear spin states where $I_a$ and $I_b$ are the nuclear spins of the atoms $a$ and $b$ and add vectorially such that their sum $I$ runs between $|I_a - I_b|$ and $|I_a + I_b|$. Because the nuclei are different, no nuclear exchange symmetry constraints are placed on the total molecular wave function. In homonuclear diatomic species, the identical nuclei do place nuclear exchange symmetry constraints on the allowed symmetry of the total molecular wavefunction (see Hougen and Oka, 2005). For the system of identical $I = 3/2$ chlorine nuclei of $Cl_2$, where both dominant isotopes $^{35}Cl$ and $^{37}Cl$ exhibit nuclear spin $I = 3/2$, in the totally symmetric ground electronic state $^1\Sigma_g^+$, there are a total of $(2I + 1)I = 6$ antisymmetric and $(2I + 1)(I + 1) = 10$ symmetric wavefunctions which must combine with rotational level wavefunctions that are even ($J = 0, 2, 4, \ldots$) and odd ($J = 1, 3, 5, \ldots$) so that the total molecular wavefunction remains antisymmetric under exchange of the fermion nuclei of the chlorine atoms. The sum over rotational levels in the previous equation then is expressed as two sums for $J$ even ($g_{\mathrm{odd}}^{\mathrm{ns}} = 6$) and $J$ odd ($g_{\mathrm{even}}^{\mathrm{ns}} = 10$), and because each rotational level of increasing energy alternates between even and odd symmetry, in the integral approximation of the rotational partition function each sum over even and odd rotational levels contributes equally resulting in a total nuclear spin degeneracy factor of $g_I^{\mathrm{ns}} = 16$ for the $Cl_2$ molecule.

For nonlinear molecules the rotational partition function becomes

$$Q_{\mathrm{rot}}^{\mathrm{nonlin}} = \frac{\boldsymbol{g}_{\boldsymbol{I}}^{\mathbf{ns}}}{\sigma}\left(\frac{k_B T}{hc}\right)^{3/2}\left(\frac{\pi}{\tilde{A}\tilde{B}\tilde{C}}\right)^{1/2} \tag{7}$$

in the integral approximation for rotational constants $\tilde{A}, \tilde{B},$ and $\tilde{C}$ corresponding to the three calculated moments of inertia and symmetry number $\sigma$. Because the rotational temperatures are much lower ($\theta_r < 0.1$ K) than the temperature range ($T = 10 - 800$ K) for all reactant species involved in reactions exhibiting barriers in this study, the integration of the rotational partition function is justified. The rotational symmetry number $\sigma$ for nonlinear molecules depends on the point group of each species; in this study, all nonlinear molecules lacking symmetry have rotational symmetry number $\sigma = 1$, and $SCl_2$, which belongs to the $C_{2v}$ point group has $\sigma = 2$. Additionally, each nonlinear species has a nuclear spin degeneracy $g_I^{\text{ns}} = \prod_j (2I_j + 1)$ contribution to the total molecular partition function. The $SCl_2$ molecule differs from the other nonlinear species in this study by having nuclear spin representation $\Gamma_{\text{ns}}^{\text{tot}} = 10A_1 \otimes 6\,B_2$ in the $C_{2v}$ point group resulting in nuclear exchange symmetry considerations for the $g_I^{\text{ns}}$ = 16 states. When coupled to states in an asymmetric top basis, each state of increasing energy exhibits alternating symmetric/antisymmetric character, and in the high temperature limit where all states contribute equally, we use $g_I^{\text{ns}}$ = 16 for the nuclear spin statistical weight to the total molecular partition function.

The partition function for each normal mode of vibration with frequency $\tilde{\nu}$ for the quantum mechanical harmonic oscillator can be written in terms of the vibration temperature $\theta_v = \frac{hc\tilde{\nu}}{k_B}$ and evaluated as a geometric series in vibrational quantum number v

$$q_v = \sum_J e^{-(hc\tilde{\nu}/k_B T)\text{v}} = \frac{1}{1-e^{-\theta_v/T}} \tag{8}$$

where the total vibrational contribution to the molecular partition function is the product from the normal modes *i*

$$Q_{\text{vib}} = \prod_i \frac{1}{1-e^{-\theta_i/T}}\,. \tag{9}$$

Many of the vibrational temperatures of the normal modes of reactants and transition states considered in this work fall within the temperature range of the calculations, suggesting the significance of their vibrational contribution to the molecular partition functions as the temperature increases from 10 to 800 K.

The molecular partition function ratio $Q = \frac{Q^{\ddagger}}{Q_A Q_B}$ can be factorized approximately as

$$Q = \prod_i \left(\frac{Q^{\ddagger}}{Q_A Q_B}\right)_i = \prod_i Q_i \tag{10}$$

into the product of the ratios of the contributions from each type of separable motion—rotation (including nuclear spin), vibration, translation, and electronic—to observe the effect each type of contribution has on the total molecular partition function ratio. We assume all reactant and transition state molecules populate only the ground electronic states, and the electronic partition function ratio $Q_{\text{elec}}$ reduces to the ratio of the degeneracies of the lowest energy ground electronic states, excluding fine structure. Because atomic species have no rotational structure and therefore no rotational contribution to the molecular partition function, we account for the nuclear spin degeneracy of the atomic species $g_I^{\text{ns}}$ as an explicit factor in the total atomic partition functions. The temperature dependences of the other partition function ratios $Q_i$ depend on the number and type of degrees of freedom available to the reactants and transition state and contributions to the average energy of the system. The temperature-dependent reaction rate coefficient obeys Arrhenius-like behavior only when $Q \propto T^{-1}$, in which case the pre-exponential factor is $\frac{k_B T}{h}\frac{Q^{\ddagger}}{Q_A Q_B} \propto T^0$ and independent of temperature.

Finally, we apply a tunneling correction factor

$$\Gamma \approx 1 - \frac{(h\nu_i)^2}{24(k_\mathrm{B}T)^2} + \,\dots\,, \tag{11}$$

in terms of the temperature $T$ and imaginary harmonic frequency $\nu_i$ of the transition state structure to reaction rate coefficients for R1–R5, those reactions with barriers along the minimum energy path (Woon and Herbst, 1996). The value of the correction factor is always greater than unity, competes with the Boltzmann factor of the transition state theory reaction rate coefficient, and can deviate far above unity at low temperatures (Bell, 1958).

*2.2.2. Exothermic reactions without a barrier*

In the absence of a barrier along the reaction coordinate, the collision between two neutral species occurs as a result of a general effective potential

$$V_\mathrm{eff}(r) = \frac{L^2}{2\mu r^2} - \frac{C_n}{r^n} \tag{12}$$

where some long-range attraction competes with the orbital angular momentum associated with the reactant systems. By defining the total energy $E$ to be greater than or equal to the maximum in the effective potential energy, the maximum impact parameter $b_\mathrm{max}$ and orbiting condition are found, and the geometric reaction cross-section

$$\sigma_r = \pi b_{max}^2 = \frac{\pi n}{n-2}\left[\frac{\boldsymbol{C_n}(n-2)}{2E}\right]^{2/n} \tag{13}$$

is written in terms of $b_\mathrm{max}$ and describes the likelihood of a reaction event in terms of the power and coefficient of the attractive potential $n$ and $C$, respectively, and the energy of the system $E$. Integrating the reaction cross section $\sigma_r$ over a Maxwell-Boltzmann distribution of energies produces the temperature-dependent, thermally averaged reaction rate coefficient

$$k(n,T)_{\mathrm{C}} = \Gamma\left(\frac{2(n-1)}{n}\right)\sqrt{\frac{8k_BT}{\pi\mu}}\,(k_BT)^{-2/n}\left[\frac{\boldsymbol{C}_{\boldsymbol{n}}(n-2)}{2}\right]^{2/n} \tag{14}$$

in terms of the attractive potential $-C/r^n$.

At long range, the induced dipoles of two neutral species obey a Lennard-Jones attraction, which varies inversely as the sixth power ($n$ = 6) of the distance between the reactant species. The coefficient of the Lennard-Jones attractive potential

$$C_{\mathrm{LJ}} = \frac{3}{2}\frac{I_A I_B}{I_A + I_B}\alpha_A\alpha_B \tag{15}$$

depends on the ionization energies $I$ and isotropic dipole polarizabilities $\alpha$ of the neutral reactants, while the induction term, which also varies inversely as the sixth power of the internuclear distance, has the coefficient

$$C_{\mathrm{dip-ind}} = \mu_a^2\alpha_b + \mu_b^2\alpha_a\,, \tag{16}$$

which depends on the permanent dipole moment $\mu$ and $\alpha$ values of the reactants (Clary et al. 1994; Herbst and Woon 1997). The general temperature-dependent reaction rate coefficient for the potential consisting of these two interactions becomes

$$k_6(T) = \Gamma(5)\sqrt{\frac{8k_BT}{\pi\mu}}\,(k_BT)^{-1/3}[2C_6]^{1/3} \tag{17}$$

where $C_6 = C_{\mathrm{LJ}} + C_{\mathrm{dip-ind}}$ and $k_6(T) \propto T^{1/6}C_6^{1/3}\mu^{-1/2}$ (Georgievskii and Klippenstein, 2005). This calculation neglects the complex structure of the identified minimum energy paths as well as nuclear spin statistics and approximates the reaction rate coefficients as a function of temperature for reactions that proceed on every collision with maximum impact parameter determined by the orbiting condition. Other terms, namely the dipole-induced-dipole and dispersion terms in the potential yield reaction rate coefficients that are generally smaller than those governed by charge-dipole and dipole-dipole interactions (Georgievskii and Klippenstein, 2005), and the capture theory reaction rate coefficients for neutral-neutral species depends more strongly on the physical conditions of the reactant system than for ion-molecule reactions (Smith et al., 2004).

## 3. Reactions involving SCl, $SCl_2$, and HSCl

### *3.1. Reactions treated and inclusion in current models*

In this work we have characterized nine reactions involving S and Cl compounds that are listed in Table 1 and designated R1–R9. All of these reactions are exothermic. The table also indicates which of the published model studies have included each reaction and what rate coefficients were used in those cases. Available experimental data are also shown (Baulch, 1981; Clyne and Townsend, 1975; Nesbitt and Leone, 1980; Fenter and Anderson, 1991; Hildebrandt et al., 1984; Murells, 1987; Sung and Setser, 1978 reported a comparable relative rate for R4). R1, R2, and R5 are abstraction reactions between $Cl_2$ and S, SH, and SCl, respectively, and have been included in various model studies as indicated in the table; rates have been measured for each of them at least at 298 K. R8 and R9 are simple barrierless addition-elimination reactions between S and SCl (R8) and two SCl molecules (R9) that have also been included in published model studies; in each case, rate coefficients were estimated based on reaction data from Murrells (1987) for other reactions that are expected to be comparable to R8 and R9. The reaction pairs R3/R4 and R6/R7 have not been included in any published model study but represent plausible reactions between $SCl_2$ and atomic H (R3/R4) or between HSCl and the OH radical (R6/R7). We will examine each of these nine reactions in greater detail in the next section using the computed reaction pathways. The connections between R1–R9 are displayed pictorially in Fig. 1, which also includes some of the endothermic reverse reactions.

When open-shell atoms and molecules react, a number of potential energy surfaces are formed corresponding to each of the possible total electronic spin states determined by vector addition of the reactant electronic spins. For reaction R1, the $^3P$ state of the sulfur atom combines with the $^1\Sigma_g^+$ chlorine molecule $Cl_2$ to form a single triplet PES. Similarly, in R2 – R7 the reactant species combine to form a single doublet state. In R8, however, the combination of the $^3P$ state of the sulfur atom and the $^2\Pi$ state of the SCl molecule results in doublet and quartet terms, so we apply a factor of 1/3 corresponding to the doublet state of the reactant complex identified in this study. The combination of the two $^2\Pi$ states of the two SCl reactants results in possible singlet and triplet states, so we apply a statistical reduction factor of 1/4 corresponding to the singlet state of the reactant complex identified in this study to the reaction rate coefficients of R9.

Table 2 tabulates various molecular properties needed to compute kinetic rate constants for the reactions, including rotational constants, dipole moments, dipole polarizabilities, ionization energies, and vibrational frequencies. These are listed as needed for reactants, transition states, and intermediates.

*3.2. Reaction outcomes and energetics*

Minimum energy pathways for R1–R9 are depicted in Fig. 2 as calculated at the RCCSD(T)/AVTZ level, with RCCSD(D)/CBS-sp values for positive barrier heights. The abstraction reactions R1, R2, and R5 each have reaction barriers that impede the process. The barriers all exceed 1.3 kcal/mol, which is large enough to yield fairly slow reaction rates at Venus cloud temperatures, as we will see below. The formation of SCl from S reacting with $Cl_2$ is the most favorable of the three reactions, with a 1.37 kcal/mol barrier. Formation of HSCl due to the SH + $Cl_2$ reaction has a barrier of 2.23 kcal/mol, and the formation of $SCl_2$ from SCl + $Cl_2$ has a barrier of 3.36 kcal/mol. SCl can be destroyed in R8 by reacting with S to yield $S_2$ and Cl, a very exothermic process that involves formation and subsequent decomposition of the SSCl chlorodisulfanyl radical complex. This intermediate is the product when two SCl radicals react in R9, which involves the disulfur dichloride intermediate, ClSSCl. R9 is much less exothermic than R8, but neither reaction is hindered by an energetic barrier.

We have also characterized two pairs of reactions in which abstraction and addition-elimination may compete with each other. In R3, H adds to $SCl_2$ to yield intermediate $HSCl_2$ after passing over a barrier of 1.61 kcal/mol. One of the Cl atoms can be eliminated exothermically to yield HSCl. This formation pathway for HSCl is expected to be much more favorable than R2, which has a substantially larger barrier height. Atomic H could also abstract a Cl atom from $SCl_2$ to yield HCl and SCl in R4, but the reaction barrier for this is 4.46 kcal/mol. Abstraction is much less favorable than addition-elimination for the H + $SCl_2$ reaction, in spite of the enormous exothermicity of over 40 kcal/mol for R4. In the other pair, the OH + HSCl reactions, the energetics are favorable for both abstraction and addition-elimination, so both pathways may yield (different) products in Venus's atmosphere. In the abstraction reaction, R6, OH removes the H atom of HSCl to yield $H_2O$ and SCl. There is an extremely small barrier of 0.04 kcal/mol. A transition state that lies below the reactant asymptote is termed a *submerged barrier* and is far less of an impediment to reaction than a positive barrier. We also found that OH can add to HSCl via a barrier that is submerged even further, at −1.99 kcal/mol, to yield the intermediate HSClOH. This species has two isomers of comparable energy that are denoted here as "a" and "b". The initial isomer formed is HSClOH-a. As the reaction diagram shows, the barrier for rearrangement between the two isomers is very low, so they can interchange easily. Once the second isomer, HSClOH-b, is formed, it can then eject the Cl atom to yield HSOH, a sulfur analog of hydrogen peroxide that is not presently included in any of the published model studies. Again, we note that abstraction is much more exothermic than addition-elimination (−35.32 kcal/mol vs −5.76 kcal/mol), but the height and character of the barriers are more critical for determining rate coefficients than reaction energies (For reactions described above with positive barriers where RCCSD(T)/CBS-sp limits are given, the full set of basis set results are provided in the Supplemental Information.)

To the best of our knowledge, only reactions R2 and R2–D have been previously characterized quantum chemically. Wang et al (2005) treated both reactions with a combination of modest Pople 6-311G(d,p) and 6-311+G(2df,2pd) basis sets in conjunction with density functional theory at the MPW1K level and molecular orbital theory at the MP2, QCISD(T), and MC-QCISD levels. Their barrier heights for reaction R2 cover a large range: MP2, 13.5 kcal/mol; MPW1K, 3.0 kcal/mol; MC-QCISD//MPW1K, 1.1 kcal/mol; and a-QCISD(T)//MP2, 0.9 kcal/mol. They found the best agreement with the experimental rate coefficients with a low value of the barrier height (their MC-QCISD//MPW1K result). Our value for the barrier height of R2–D is 1.94 kcal mol$^{-1}$.

The diagrams in Fig. 2 also indicate the energetics for the reverse reactions of R1-R9. They all have barriers of 3.8 kcal/mol or larger (and often *much* larger), so they are expected to have very small rate coefficients in the upper and middle atmospheres of Venus and will not be treated further in this study. One point to note is that some of the intermediates for either forward or reverse reactions could potentially be stabilized by non–reactive collisions with other molecules in Venus's atmosphere, mostly $CO_2$, which could possibly yield species incluiding $HSCl_2$ (R3), HSClOH(a,b) (R7), SSCl (R8), and ClSSCl (R9).

*3.3. Partition function contributions and competition*

Fig. 3 shows the temperature-dependent partition function ratios $Q_i(T)$ for each type of motion—translation, rotational, and vibrational—and for the total molecular partition function for each of the reactions R1–R5, each of which exhibits a barrier along the minimum energy reaction coordinate. The panels in Figure 3 show the competing effects of each contribution to the total partition function ratio as a function of temperature as well as the temperature-dependent contributions of the exponential and tunneling correction factors to the overall reaction rate coefficients to illustrate which factors dominate at which temperatures. For reaction pathways with identical reactants (*e.g.*, R3 and R4) or very similar reactants (*e.g*,. R1, R2, and R2–D), the translational (R3/R4 and R1/R2/R2-D) and rotational (R3/R4 and R2/R2–D) partition function ratios lie closely and parallel to one another (see Fig. 3, top left and middle left). All translational partition function ratios exhibit the same temperature dependence, $Q_{\mathrm{trans}} \propto T^{-3/2}$, reflecting the three translational degrees of freedom available to each reactant and transition state. The temperature dependence of the rotational contribution to the total partition function ratio depends on the geometry of the reactant species and transition states. The rotational partition function ratio for R1 between atomic S and homonuclear diatomic $Cl_2$ is $Q_{\mathrm{rot}} \propto T^{1/2}$, while the rotational partition function ratio for R2 and R5 between heteronuclear diatomic molecules SH and SCl and homonuclear diatomic $Cl_2$ is $Q_{\mathrm{rot}} \propto T^{-1/2}$. Because the sum of the rotational degrees of freedom of the reactant species H and $SCl_2$ are equal to the number of rotational degrees of freedom of the transition state in R3 and R4, the total rotational partition function ratios $Q_{\mathrm{rot}}$ remain constant as a function of temperature. All vibrational partition function ratios increase over the temperature range for R1–R5 (see Fig. 3, top right), and this trend appears as a result of the lower equivalent vibrational temperatures of the normal modes of the transition states compared with the equivalent vibrational temperatures of the normal modes of the reactant species. Minima in the total partition function ratios $Q_{\mathrm{total}}$ (see Fig. 3, top center) appear where the ratios of the transition state partition function and the product of the reactant partition functions to their respective first derivatives are equal and opposite. Reactions R3 and R4, which share common reactants (H + $SCl_2$), exhibit very similar translational and rotational partition function ratios (Fig. 3, top left and top center), but a greater difference between the vibrational partition function ratios for these reactions solely accounts for the larger difference in the total partition function ratios for these reactions. Fig. 3 also shows the logarithm of the tunneling correction factor (middle right) and temperature-dependent reaction rate coefficients (bottom left) for reactions R1–R5.

*3.4. Temperature-dependent rate coefficients for reactions with barriers*

For reactions with barriers (R1–R5), Fig. 4 shows the calculated temperature-dependent transition state theory reaction rate coefficients, $k_{\mathrm{TST}}(T)$, as dotted lines with color that corresponds to the reaction number.

Chemical kinetic models of the Venusian atmosphere utilize the Arrhenius-Kooij formula

$$k(T) = \alpha\left(\frac{T}{300\ \mathrm{K}}\right)^{\beta} e^{-\gamma/T} \tag{18}$$

for tabulating reaction rate coefficient data, so we fit the Arrhenius-Kooij parameters $\alpha$ (not to be confused with the dipole polarizability) and $\beta$ to our calculations so these reactions can be included in future kinetic models of the Venusian atmosphere**.** We determine the parameters $\alpha$ and $\beta$ by minimizing the sum of the squares of the residuals between the calculated $k_{\mathrm{TST}}(T)$ pre-exponential factors and the Arrhenius-Kooij reaction rate coefficient pre-exponential factors $\ln(A) = \ln(\alpha) + \beta \ln(T/300\ \mathrm{K})$, while $\gamma$ has been determined from the electronic structure calculations and is excluded from the least-squares fitting procedure. Table 3 contains the Arrhenius-Kooij parameters $\alpha$, $\beta$, and $\gamma$ as well as the coefficient of determination $R^2$ for each reaction, and these optimized $k_{\mathrm{AK}}(T)$ appear as solid lines in Fig. 4.

Because the partition functions and partition function ratios do not exhibit a simple power-law dependence on the temperature due to the vibrational contribution at low temperatures $T < \Theta_v$, the pre-exponential factors of R1–R5 are not well fit by the power law expression of the Arrhenius-Kooij formula over the entire temperature range (see Table 3). To find better fit representations for atmospheric kinetic models, we fit Arrhenius-Kooij parameters $\alpha$ and $\beta$ to the calculated reaction rate coefficients $k_{\mathrm{TST}}(T)$ piecewise (piecewise Arrhenius-Kooij, or PAK in our notation) over three reduced temperature ranges, low ($T <$ 50 K, type 1), medium (50 K < $T$ < 150 K, type 2), and high ($T >$ 150 K, type 3), and record them in Table 3. Substantial increases in the coefficients of determination $R^2$ emerge for $k_{\mathrm{PAK}}(T)$ when compared with those $R^2$ for $k_{\mathrm{AK}}(T)$, indicating better fits over the piecewise temperature ranges, the boundaries of which appear in black solid vertical lines in Fig. 4. These boundaries are motivated by the temperature ranges of models of the lower and middle atmosphere of Venus, and in Fig. 4, these modeled temperature ranges appear as different background colors (see Fig. 4 caption for references).

Reaction R1 between atomic sulfur and diatomic chlorine, $S + Cl_2 \rightarrow Cl + SCl$, is represented in several models of the middle and lower atmosphere of Venus (Mills and Allen, 2007; Krasnopolsky, 2007; Krasnopolsky, 2012; Zhang, 2012) with an Arrhenius equation $k(T) = 2.8 \times 10^{-11}\, e^{-290/T}$ $\mathrm{cm^3 s^{-1}}$. Our calculated rate coefficients $k_{\mathrm{TST}}(T)$ for R1 are several orders of magnitude smaller than the existing Arrhenius values for all temperatures and exhibit non-Arrhenius behavior as a result of the lack of a simple power-law temperature dependence of the vibrational partition functions. The large discrepancy between the older value and ours is due to large differences in both the pre-exponential factor and the exponent. Our calculated reaction rate coefficients for the reaction between SH and $Cl_2$, R2, are several orders of magnitude lower than the literature value $k(T) = 1.4 \times 10^{-11}\, e^{-690/T}$ $\mathrm{cm^3 s^{-1}}$ (Sander et al., 2006; Krasnopolsky, 2012), which also exhibits slightly less variation over middle atmosphere temperatures, for all temperatures in our range. Once again, the large discrepancy between the older value and ours is due both to differences in the pre-exponential and exponential factors. The calculated reaction rate coefficients for R3 and R4, reactions between H and $SCl_2$, are roughly within an order of magnitude of each other, $\log(k_{R3}) \sim \log(k_{R4})$, at higher temperatures $T$ > 300 K but quickly diminish and diverge as the temperature falls below 300 K. Because reactions R3 and R4 are absent in existing chemical kinetic models of the lower and middle atmosphere of Venus, our calculated values provide new constraints on the coupled kinetics of sulfur and chlorine chemistry in atmospheric models. At temperatures below 800 K, our calculated temperature-dependent reaction rate coefficients for R5, the reaction between SCl and $Cl_2$, fall several orders of magnitude below the reported temperature-independent value $k = 7 \times 10^{-14}\ \mathrm{cm^3 s^{-1}}$ (Murrells, 1987), which has been incorporated in chemical models of the middle atmosphere (~50–112 km) of Venus (Mills and Allen, 2007; Krasnopolsky, 2012), and this difference could impact existing chemical models of the middle atmosphere due to the decreased rate coefficient values at the middle atmosphere temperatures ($T$ ~150–375 K).

For reactions R1 and R2-D, the experimental uncertainties presented in Table 1 are reflected in Fig. 4 as shaded regions surrounding the dot-dashed curves expressing the measured temperature-dependent rate coefficients. For the rate coefficients measured at a single temperature, 298 K, for reactions R1, R2, and R4, the errors are so small that they appear within the "•" marks on the scale of Fig. 4. To estimate the uncertainties in our calculated temperature-dependent reaction rate coefficients for reactions R1 and R2-D, those reactions with measured barriers, we calculate half of the differences in the measured and calculated barriers (199.5 K for R1 and 143 K for R2-D) and shade the region surrounding the calculated values (dotted lines near dashed lines in Fig. 4).

*3.5. Rate coefficients for reactions without barriers*

Fig. 5 shows the temperature-dependent capture theory reaction rate coefficients $k_C(T)$ for the reactions without barriers, R6–R9**.** All rate coefficients vary with a power law temperature dependence $k(T) \propto T^{1/6}$ as a result of the Lennard-Jones attraction and induction terms included in the effective potential. Thus, in contrast to the TST calculations, these rate coefficients do not warrant a least-squares method to fit the Arrhenius-Kooij formula for implementation into existing chemical networks for kinetic models of the Venusian atmosphere**.** Because these reaction rate coefficients depend only on the reactant electronic structures, R6 and R7 exhibit identical temperature-dependent reaction rate coefficients $k_C(T)$ as they increase slightly over the temperature range $T$ = 10–800 K. The reactions R6 and R7 do not appear in current chemical models of the lower and middle Venusian atmosphere, and our calculations provide data for the rate coefficients of the reactions OH + HSCl → SCl + $H_2O$ (R6) and OH + HSCl → Cl + HSOH (R7) that can be included in existing chemical networks. Some chemical models of the lower and middle atmosphere of Venus (Mills and Allen, 2007; Zhang, 2012) include reaction R8, S + SCl → Cl + $S_2$, in their chemical network with a constant rate coefficient $k(295\ \mathrm{K}) = 1 \times 10^{-11}\ \mathrm{cm^3 s^{-1}}$ (see Fig. **5** black "•", Murrells, 1987) and $k(295\ \mathrm{K}) = 1 \times 10^{-12}\ \mathrm{cm^3 s^{-1}}$ used in an earlier model (Krasnopolsky, 2007), slightly lower than our calculated temperature dependent reaction rate coefficients. The reaction between two sulfur monochloride molecules, SCl + SCl → Cl + SSCl (R9), is present in chemical networks used to model the middle atmosphere of Venus (Miller and Allen, 2007; Zhang, 2012) with a constant rate coefficient $k(295\ \mathrm{K}) = 5.4 \times 10^{-11}\ \mathrm{cm^3 s^{-1}}$ (Murrells, 1987), and this value is slightly greater than our calculated values. Our calculated reaction rate coefficients, however, are temperature-dependent, albeit weakly, and can be used in chemical networks to model the chemistry of the Venusian atmosphere over the appropriate temperature ranges. For reactions R6 and R7, which have identical reactant and different product channels, unknown branching ratios can be estimated by comparison to reactions with known fractions and similar reactants, and the rate coefficients reported in this study can be scaled by the appropriate fractions for each unique product channel.

**4. Conclusions**

Using the updated structural parameters of reactant species and identified transition states, we have calculated the temperature-dependent reaction rate coefficients for a set of nine reactions of possible importance in kinetic models of the atmosphere of Venus. For reactions R1–R5 with identified barriers along the minimum energy path, we calculated the transition state theory reaction rate coefficients $k_{\mathrm{TST}}(T)$, corrected for tunneling, and nuclear spin statistics when calculating the molecular partition functions when appropriate. For reactions without barriers, R6–R9, we have calculated the reaction rate coefficients $k_{\mathrm{C}}(T)$ using a capture model and an effective potential including terms for orbital angular momentum, a Lennard-Jones attraction, and dipole-induced dispersion. To provide data that can be integrated into existing chemical networks and kinetic models of various layers of the atmosphere of Venus, we fit Arrhenius-Kooij parameters to reaction rate coefficients $k_{\mathrm{PAK}}(T)$ for reactions with barriers over three temperature ranges characteristic of a low temperature limit and the middle and lower Venusian atmospheres. Our calculated reaction rate coefficients are generally lower than adopted values in existing models of the Venusian atmosphere. Expanding the kinetic model to consider Venus-like pressures and possible collisional stabilization will, in the future**,** provide stronger constraints on the reaction rate coefficients for the nine systems in this study. Additionally, our study can supplement chemical networks with reactions involving deuterated species, which can guide observations of these isotopologues in future molecular line surveys**.** The Arrhenius-Kooij parameters we report for each of the reactions reflect the updated geometries and new transition state energies identified in this study and can be readily incorporated into existing chemical networks and compared with future experimental and theoretical efforts over a broad temperature range.

**Acknowledgments**

We gratefully acknowledge the support of the NASA Planetary Atmospheres program through Grant NNX14AK32G. We also acknowledge helpful conversations with Dr. Hua Guo (University of New Mexico) concerning the kinetic calculations.

**Supplementary materials.**

Supplementary data to this article can be found online at https://doi.org/10.1016/j.icarus.2020.114051.

**Table 1**

Reactions characterized in this work.

| Reaction | | Model (Reaction) | Rate Coefficient ($cm^3\ s^{-1}$) | Source |
|---|---|---|---|---|
| R1 | $S + Cl_2 \rightarrow Cl + SCl$ | Mills and Allen, 2007 (R266)<br>Krasnopolsky, 2007 (R55)<br>Krasnopolsky, 2012 (R139)<br>Zhang et al., 2012 (R180) | $(2.8\pm0.1) \times 10^{-11}\ e^{(-290\pm60)/T}$<br>$(1.1\pm0.1) \times 10^{-11}$ (298 K) | Baulch, 1981<br>Clyne and Townsend, 1975 |
| R2 | $SH + Cl_2 \rightarrow Cl + HSCl$ | Krasnopolsky, 2007 (R59) | $(1.36\pm0.2) \times 10^{-12}$ (298 K) | Nesbitt and Leone, 1980 |
| R2-D | $SD + Cl_2 \rightarrow Cl + DSCl$ | … | $(1.7\pm0.4) \times 10^{-11}\ e^{(-690\pm90)/T}$ | Fenter and Anderson, 1991 |
| R3 | $H + SCl_2 \rightarrow Cl + HSCl$ | … | … | … |
| R4 | $H + SCl_2 \rightarrow SCl + HCl$ | … | $(0.71\pm0.03)\ k_{H+Cl_2}$<br>$k_{H+Cl_2}$: $2.0 \times 10^{-11}$ (298 K) | Hildebrandt et al., 1984<br>Berho et al., 1999 |
| R5 | $SCl + Cl_2 \rightarrow Cl + SCl_2$ | Mills and Allen, 2007 (R337)<br>Krasnopolsky, 2012 (R142) | $7 \times 10^{-14}$ | Murrells, 1987 |
| R6 | $OH + HSCl \rightarrow SCl + H_2O$ | … | … | … |
| R7 | $OH + HSCl \rightarrow Cl + HSOH$ | … | … | … |
| R8 | $S + SCl \rightarrow Cl + S_2$ | Mills and Allen, 2007 (R336)<br>Krasnopolsky, 2007 (R57)<br>Zhang et al., 2012 (R272) | $1 \times 10^{-11}$ | Murrells, 1987 (est) |
| R9 | $SCl + SCl \rightarrow Cl + SSCl$ | Mills and Allen, 2007 (R339)<br>Zhang et al., 2012 (R276) | $5.4 \times 10^{-11}$ | Murrells, 1987 (est) |

**Table 2**
Calculated properties of stationary points (minima and TS) for the reactions in Table 1.

| Species | Rotational Constants ($cm^{-1}$) | μ (D) | α (Å$^3$) | IE (eV) | Frequencies ($cm^{-1}$) |
|---|---|---|---|---|---|
| S | … | 0.000 | 2.144 | 10.36 | … |
| Cl | … | 0.000 | 2.865 | 12.97 | … |
| $Cl_2$ | 0.23895 | 0.000 | 4.536 | 11.48 | 546 |
| SCl | 0.25366 | 0.097 | 5.253 | 9.63[a] | 567 |
| SH | 9.5701 | 0.749 | 3.264 | 10.42 | 2700 |
| SD | 4.9351 | 0.749 | 3.264 | 10.42 | 969 |
| OH | 18.8397 | 1.639 | 1.075 | 13.02 | 3748 |
| HCl | 10.5776 | 1.075 | 2.538 | 12.74 | 3005 |
| $S_2$ | 0.29051 | 0.000 | 5.899 | 9.36 | 718 |
| HSCl | 0.23039, 0.23597, 9.73941 | 1.155 | 5.509 | 9.92[a] | 529, 922, 2704 |
| $SCl_2$ | 0.07988, 0.09597, 0.47648 | 0.401 | 7.642 | 9.47 | 205, 522, 523 |
| SSCl | 0.08071, 0.09328, 0.59897 | 0.712 | 9.265 | 10.10[a] | 212, 439, 660 |
| $H_2O$ | 9.4797, 14.5719, 27.1274 | 1.845 | 1.407 | 12.62 | 1650, 3841, 3951 |
| HSOH | 0.49278, 0.50654, 6.7501 | 1.576 | 4.207 | 9.41[a] | 479, 772, 1026, 1209, 2658, 3833 |
| ClSSCl | … | … | … | … | 92, 204, 240, 461, 471, 540 |
| HSClOH-a | … | … | … | … | 191, 307, 356, 479, 640, 1036, 1208, 2764, 3804 |
| HSClOH-b | … | … | … | … | 223, 304, 457, 543, 820, 1030, 1237, 2641, 3595 |
| TS1 | 0.05071, 0.05132, 4.2551 | … | … | … | 241i, 111, 347 |
| TS2 | 0.05049, 0.05123, 2.0531 | … | … | … | 251i, 125, 233, 349, 549, 2701 |
| TS2-D | 0.04966, 0.05022, 1.6891 | … | … | … | 251i, 124, 183, 347, 399, 1940 |
| TS3 | 0.07700, 0.09491, 0.39417 | … | … | … | 581i, 192, 199, 283, 498, 521 |
| TS4 | 0.07342, 0.08844, 0.43083 | … | … | … | 911i, 92, 171, 254, 493, 530 |
| TS5 | 0.02711, 0.02960, 0.27756 | … | … | … | 294i, 66, 140, 247, 341, 553 |
| TS6 | 0.08578, 0.11257, 0.34510 | … | … | … | 780i, 77, 173, 223, 535, 669, 1009, 1786, 3744 |
| TS7 | 0.08472, 0.12220, 0.26382 | … | … | … | 214i, 100, 112, 183, 333, 527, 838, 2702, 3734 |
| TS8 | 0.08680. 0.09484, 0.84623 | … | … | … | 145i, 254, 287, 532, 773, 1042, 1185, 2676, 3776 |

[a] RCCSD(T)/CBS value (all other IEs from the NIST Webbook; Linstrom & Mallard, 2018)

**Table 3**

Arrhenius-Kooij Parameters.

| ID | Reaction | Type | $\alpha$ ($cm^3\ s^{-1}$) | $\beta$ | $\gamma$ (K) | $R^2$ |
|---|---|---|---|---|---|---|
| R1 | $S + Cl_2 \rightarrow Cl + SCl$ | AK | $1.696\times10^{-13}$ | $1.843\times10^{-1}$ | $6.892\times10^{2}$ | 0.108 |
| | | 1 | $5.861\times10^{-15}$ | $-1.712\times10^{0}$ | $6.892\times10^{2}$ | 0.997 |
| | | 2 | $5.770\times10^{-14}$ | $-3.935\times10^{-1}$ | $6.892\times10^{2}$ | 0.803 |
| | | 3 | $1.310\times10^{-13}$ | $8.620\times10^{-1}$ | $6.892\times10^{2}$ | 1.000 |
| R2 | $SH + Cl_2 \rightarrow Cl + HSCl$ | AK | $1.498\times10^{-14}$ | $-3.554\times10^{-1}$ | $1.122\times10^{3}$ | 0.170 |
| | | 1 | $2.762\times10^{-16}$ | $-2.744\times10^{0}$ | $1.122\times10^{3}$ | 0.999 |
| | | 2 | $2.883\times10^{-15}$ | $-1.365\times10^{0}$ | $1.122\times10^{3}$ | 0.971 |
| | | 3 | $9.668\times10^{-15}$ | $7.513\times10^{-1}$ | $1.122\times10^{3}$ | 1.000 |
| R2-D | $SD + Cl_2 \rightarrow Cl + DSCl$ | AK | $6.884\times10^{-15}$ | $-2.564\times10^{-1}$ | $9.760\times10^{2}$ | 0.095 |
| | | 1 | $1.034\times10^{-16}$ | $-2.739\times10^{0}$ | $9.760\times10^{2}$ | 0.999 |
| | | 2 | $1.258\times10^{-15}$ | $-1.275\times10^{0}$ | $9.760\times10^{2}$ | 0.962 |
| | | 3 | $4.481\times10^{-15}$ | $8.380\times10^{-1}$ | $9.760\times10^{2}$ | 1.000 |
| R3 | $H + SCl_2 \rightarrow Cl + HSCl$ | AK | $2.027\times10^{-12}$ | $-7.195\times10^{-1}$ | $8.100\times10^{2}$ | 0.520 |
| | | 1 | $1.496\times10^{-13}$ | $-2.421\times10^{0}$ | $8.100\times10^{2}$ | 1.000 |
| | | 2 | $4.369\times10^{-13}$ | $-1.767\times10^{0}$ | $8.100\times10^{2}$ | 0.991 |
| | | 3 | $1.329\times10^{-12}$ | $3.204\times10^{-1}$ | $8.100\times10^{2}$ | 0.999 |
| R4 | $H + SCl_2 \rightarrow SCl + HCl$ | AK | $5.215\times10^{-12}$ | $-7.272\times10^{-1}$ | $2.244\times10^{3}$ | 0.546 |
| | | 1 | $3.632\times10^{-13}$ | $-2.424\times10^{0}$ | $2.244\times10^{3}$ | 1.000 |
| | | 2 | $1.291\times10^{-12}$ | $-1.671\times10^{0}$ | $2.244\times10^{3}$ | 0.992 |
| | | 3 | $3.479\times10^{-12}$ | $2.684\times10^{-1}$ | $2.244\times10^{3}$ | 0.999 |
| R5 | $SCl + Cl_2 \rightarrow Cl + SCl_2$ | AK | $2.485\times10^{-15}$ | $8.052\times10^{-2}$ | $1.690\times10^{3}$ | 0.009 |
| | | 1 | $2.130\times10^{-17}$ | $-2.705\times10^{0}$ | $1.690\times10^{3}$ | 0.998 |
| | | 2 | $3.932\times10^{-16}$ | $-1.008\times10^{0}$ | $1.690\times10^{3}$ | 0.926 |
| | | 3 | $1.564\times10^{-15}$ | $1.266\times10^{0}$ | $1.690\times10^{3}$ | 1.000 |
| R6 | $OH + HSCl \rightarrow SCl + H_2O$ | C | $3.886\times10^{-11}$ | $1.667\times10^{-1}$ | … | … |
| R7 | $OH + HSCl \rightarrow Cl + HSOH$ | C | $3.886\times10^{-11}$ | $1.667\times10^{-1}$ | … | … |
| R8 | $S + SCl \rightarrow Cl + S_2$ | C | $1.175\times10^{-11}$ | $1.667\times10^{-1}$ | … | … |
| R9 | $SCl + SCl \rightarrow Cl + SSCl$ | C | $9.420\times10^{-11}$ | $1.667\times10^{-1}$ | … | … |

Note: Fit type C corresponds to the capture theory formula $k_C = \alpha(T/300\ K)^{\beta}$ where $\beta = 1/6$ for neutral-neutral reactions, fit type AK corresponds to fits of the Arrhenius-Kooij pre-exponential factor $\log(A) = \log(\alpha) + \beta \log(T/300\ K)$ for all temperatures ($10 < T < 800$), while fit types 1, 2, and 3 correspond to the fits of the piecewise Arrhenius-Kooij pre-

exponential factor $\log(A_{PAK}) = \log(\alpha) + \beta \log(T/300)$ for low ($T < 50$ K), medium (50 K $\leq T \leq$ 150 K), and high temperature ranges ($T > 150$ K).

FIGURES AND FIGURE CAPTIONS

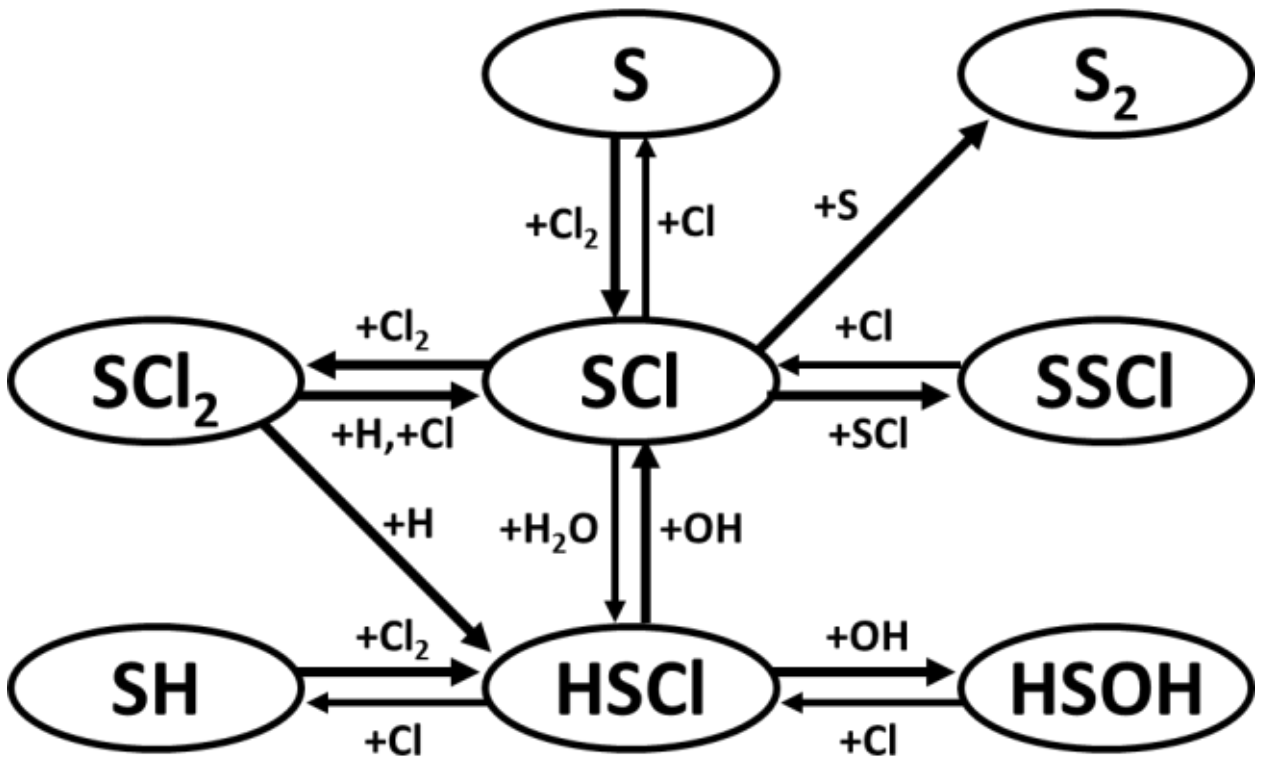

**Fig. 1**. Selected formation and destruction reactions for SCl, $SCl_2$, and HSCl of relevance to Venus atmospheric chemistry.

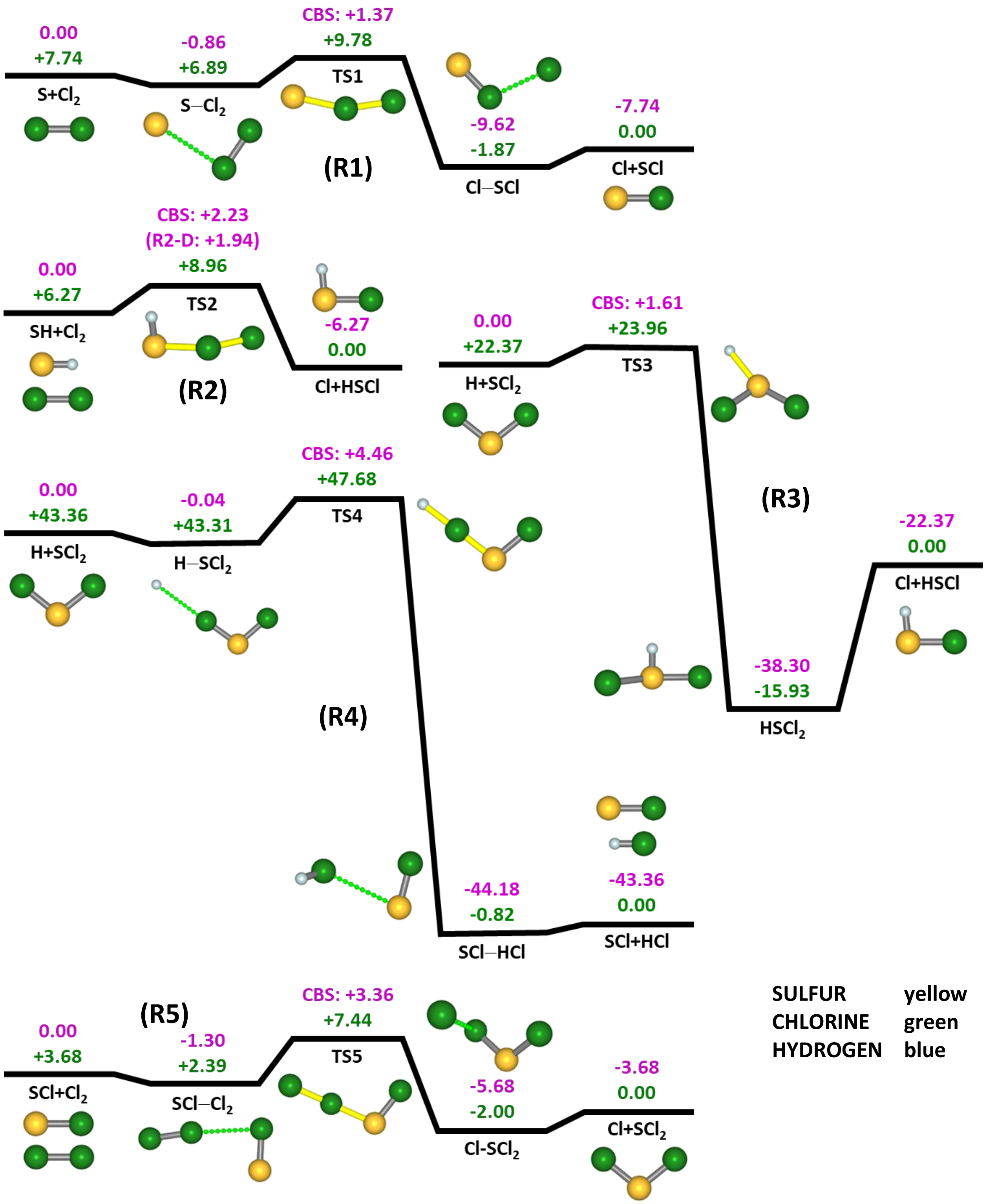

**Fig. 2.** Reaction diagrams for processes shown in Fig. 1. Energies are in kcal/mol relative to the reactant asymptote, with magenta used for forward reactions and green for reverse reactions. All values are at the RCCSD(T)/AVTZ level, except as noted [CBS indicates RCCSD(T)/CBS-sp]. The barrier height for the deuterated reaction R2-D is indicated on the diagram for R2. ZPE corrections are included.

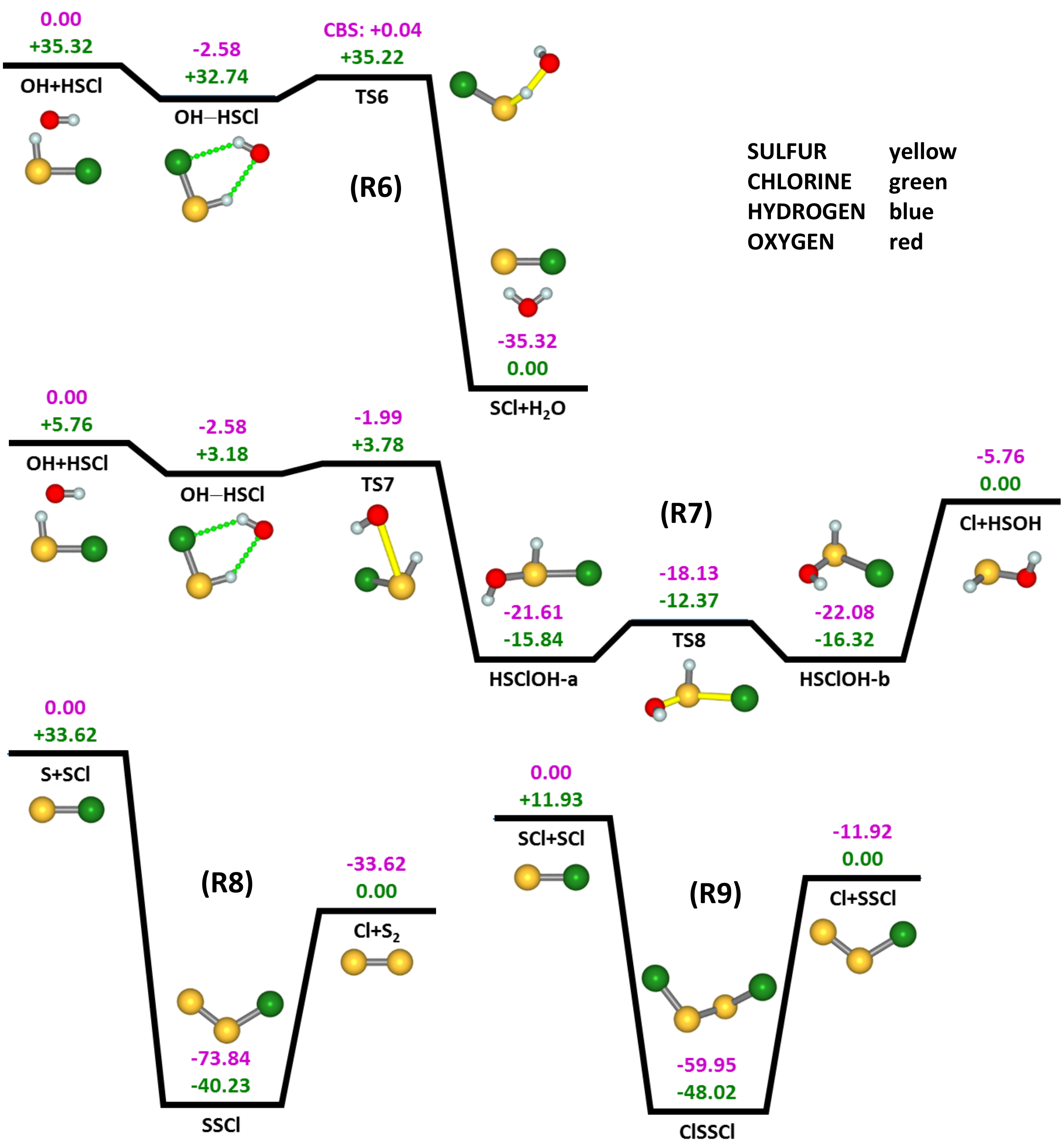

**Fig. 2.** Continued.

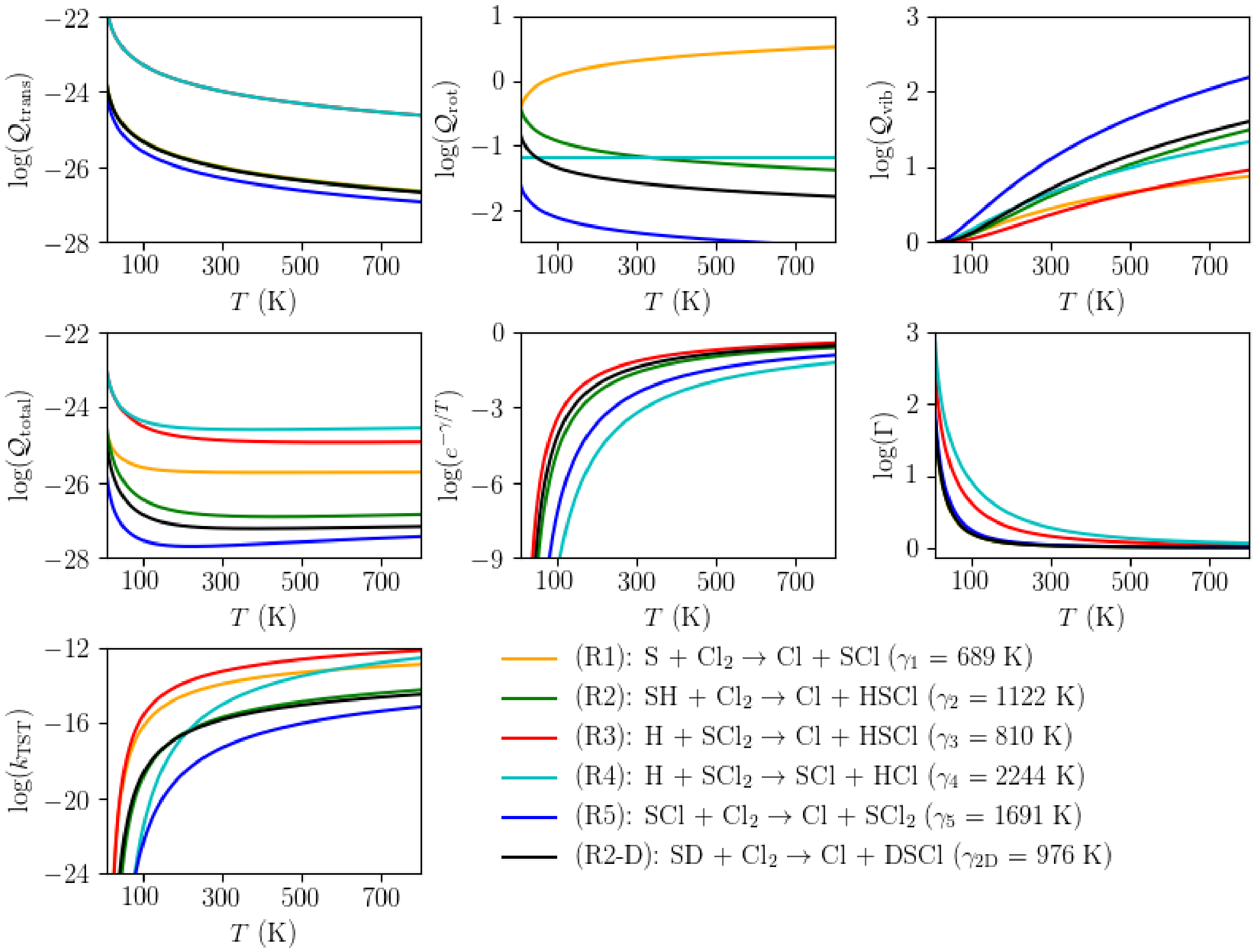

**Fig. 3.** Ratios of the logarithms of the temperature-dependent partition function contributions for various degrees of freedom: translational (top left), rotational (top center), vibrational (top right), the total molecular partition (middle left), the Boltzmann factor (middle center) the tunneling correction (middle right), and the transition-state-theory reaction rate coefficients $k_{TST}(T)$ (bottom left) for reactions R1–R5 where color corresponds to reaction number.

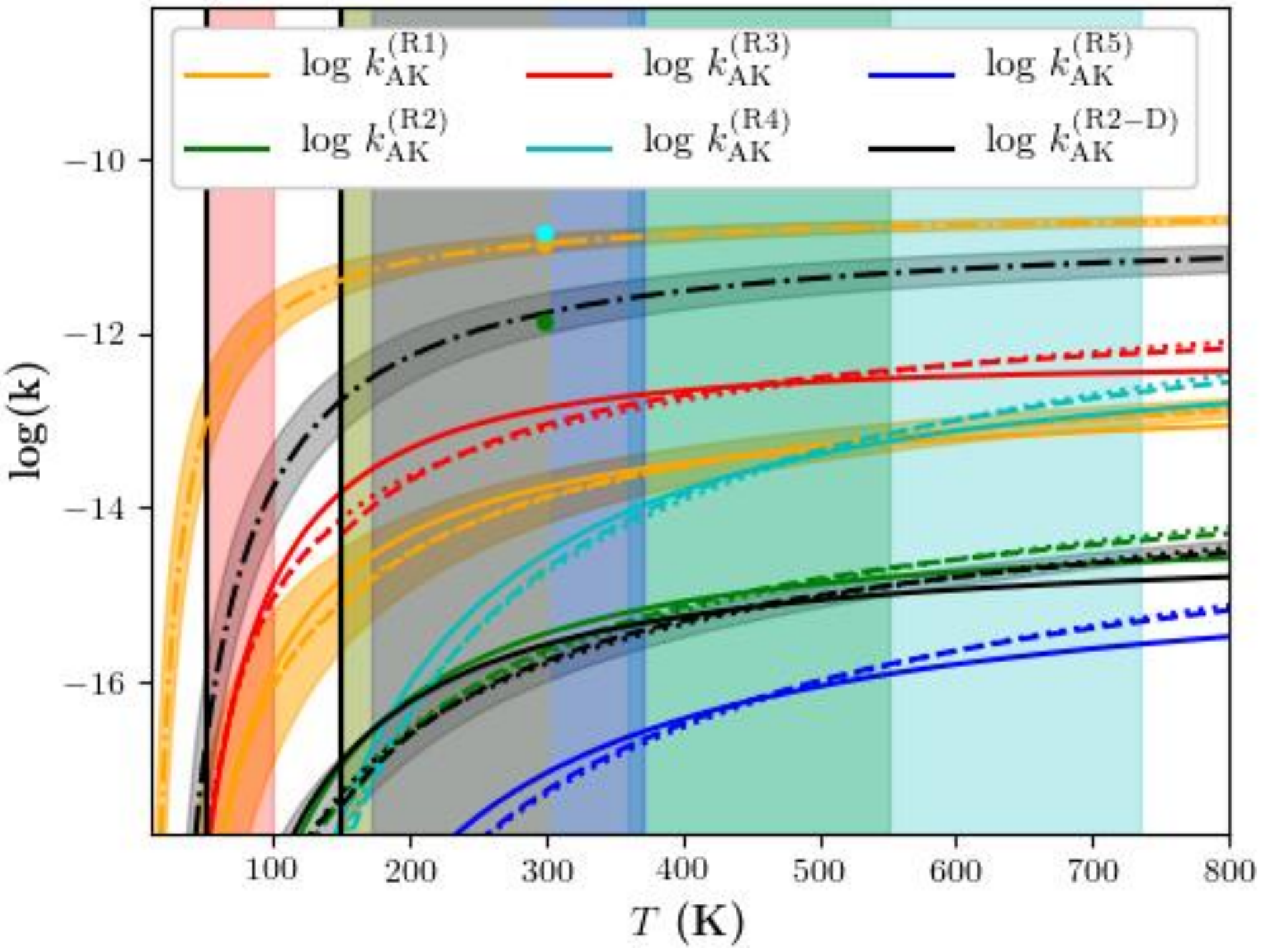

**Fig. 4.** Temperature-dependent reaction rate coefficients $k$(T) for reactions R1 – R5. Colors correspond to reaction number, while line styles correspond to formula fit by least-squares method: transition state theory calculations (TST, dotted), Arrhenius-Kooij fits to the TST calculations (AK, solid), and modified piecewise Arrhenius-Kooij fits over the temperature range fragments (PAK, dashed). Points marked with "•" refer to values determined from previous studies (see text). Solid black vertical lines appear at the temperature range fragment boundaries for the PAK fits, and background color corresponds to the temperature ranges (and altitude ranges) of atmospheric kinetic models of the Venusian atmosphere: red, 50–100 K (112–58 km); goldenrod, 150–300 K (112–58 km); green, 150–550 K (130–80 km); blue, 172.2–370.2 K (112–47 km); and cyan, 360–735 K (47–0 km) (see text for references).

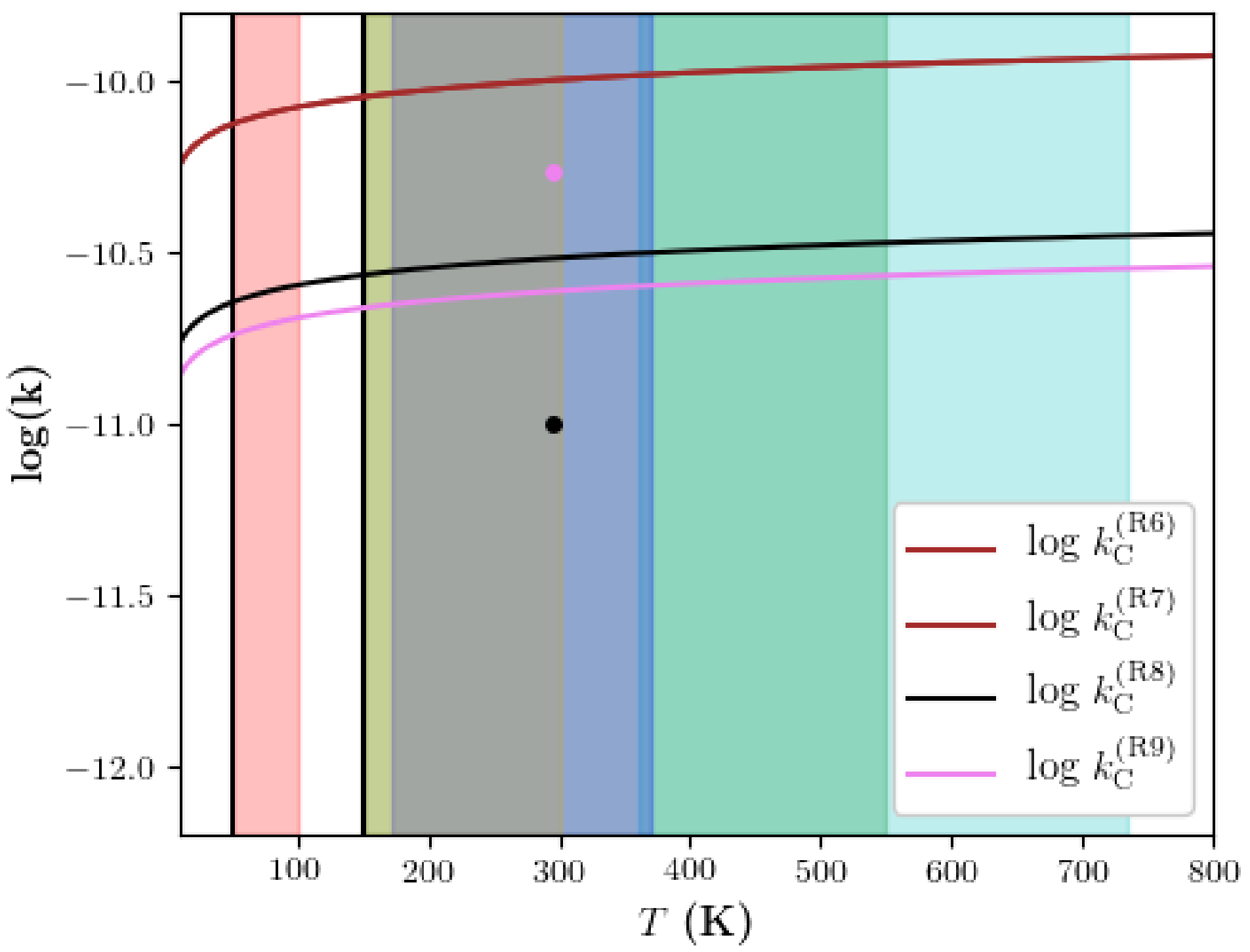

**Fig. 5.** Temperature-dependent capture theory reaction rate coefficients *k*(T) for reactions R6–R9. Colors correspond to reaction number. Points marked with "•" refer to literature values (see text); see Fig. 4 for key to background colors.

Supplemental Material for "Theoretical kinetic studies of Venus chemistry. Formation and destruction of SCl, $SCl_2$, and HSCl" by D. E. Woon, D. M. Maffucci, and E. Herbst

## A. BASIS SET RESULTS FOR REACTIONS[a]

| RXN | TZ | QZ | 5Z | CBS |
|---|---|---|---|---|
| R1 | 2.03 | 1.62 | 1.46 | 1.37 |
| R2 | 2.69 | 2.42 | 2.31 | 2.23 |
| R2-D | 2.40 | 2.13 | 2.02 | 1.94 |
| R3 | 1.59 | 1.60 | 1.61 | 1.61 |
| R4 | 4.32 | 4.41 | 4.45 | 4.46 |
| R5 | 3.76 | 3.54 | 3.43 | 3.36 |
| R6 | -0.11 | -0.12 | -0.05 | 0.04 |

[a] Barrier heights in kcal/mol; RCCSD(T) complete basis set limits (CBS) calculated from extrapolated total energies of reactants and the transition state, including zero point energy corrections.

## B. BASIS SET RESULTS FOR CALCULATED IONIZATION ENERGIES[a]

| SPECIES | TZ | QZ | 5Z | CBS |
|---|---|---|---|---|
| SH | 10.26 | 10.34 | 10.37 | 10.38 |
| SCl | 9.52 | 9.60 | 9.62 | 9.63 |
| HSCl | 9.83 | 9.90 | 9.91 | 9.92 |
| SSCl | 9.97 | 10.07 | 10.09 | 10.10 |
| HSOH | 9.34 | 9.39 | 9.41 | 9.41 |

[a] Ionization energies in ev; RCCSD(T) complete basis set limits (CBS) calculated from extrapolated total energies of neutral and cationic species; zero point energy corrections included.

## C. COORDINATES AND ENERGIES

----------------------------------------

**H($^2$S)** [H_2S.xyz]

----------------------------------------

```
RHF/AVTZ (Eh)       -0.499821
RHF/AVQZ-sp (Eh)    -0.499948
RHF/AV5Z-sp (Eh)    -0.499995
RHF/CBS-sp (Eh)     -0.500023
ZPE (kcal/mol)              0.000
Cartesian coordinates in Angstroms:
 H     0.000000     0.000000     0.000000
--------------------------------------------
```

**S($^3$P)** [S_3P.xyz]

```
--------------------------------------------
RCCSD(T)/AVTZ (Eh)       -397.656925
RCCSD(T)/AVQZ-sp (Eh)    -397.668282
RCCSD(T)/AV5Z-sp (Eh)    -397.671759
RCCSD(T)/CBS-sp (Eh)     -397.673293
ZPE (kcal/mol)              0.000
Cartesian coordinates in Angstroms:
 S     0.000000     0.000000     0.000000
--------------------------------------------
```

**Cl($^2$P)** [Cl_2P.xyz]

```
--------------------------------------------
RCCSD(T)/AVTZ (Eh)       -459.677652
RCCSD(T)/AVQZ-sp (Eh)    -459.695625
ZPE (kcal/mol)              0.000
Cartesian coordinates in Angstroms:
CL     0.000000     0.000000     0.000000
--------------------------------------------
```

**$Cl_2$($^1\Sigma_g^+$)** [Cl2_1Sigmag+.xyz]

```
--------------------------------------------
RCCSD(T)/AVTZ (Eh)       -919.442880
RCCSD(T)/AVQZ-sp (Eh)    -919.483280
RCCSD(T)/AV5Z-sp (Eh)    -919.496015
RCCSD(T)/CBS-sp (Eh)     -919.501877
ZPE (kcal/mol)              0.780
Cartesian coordinates in Angstroms:
CL     0.000000     0.000000    -1.004616
CL     0.000000     0.000000     1.004616
```

```
--------------------------------------------
```

**SCl($^2\Pi$)** [SCl_2Pi.xyz]

```
--------------------------------------------
RCCSD(T)/AVTZ (Eh)       -857.434540
RCCSD(T)/AVQZ-sp (Eh)    -857.469072
RCCSD(T)/AV5Z-sp (Eh)    -857.480025
RCCSD(T)/CBS-sp (Eh)     -857.485113
ZPE (kcal/mol)              0.810
```

```
Cartesian coordinates in Angstroms:
 S     0.000000     0.000000    -1.047798
CL     0.000000     0.000000     0.947520
--------------------------------------------
```

**SH($^2\Pi$)** [SH_2Pi.xyz]

```
--------------------------------------------
RCCSD(T)/AVTZ (Eh)       -398.293091
RCCSD(T)/AVQZ-sp (Eh)    -398.306680
RCCSD(T)/AV5Z-sp (Eh)    -398.310835
RCCSD(T)/CBS-sp (Eh)     -398.312665
ZPE (kcal/mol)              3.860
Cartesian coordinates in Angstroms:
 S     0.000000     0.000000    -0.040949
 H     0.000000     0.000000     1.302496
--------------------------------------------
```

**OH($^2\Pi$)** [OH_2Pi.xyz]

```
--------------------------------------------
RCCSD(T)/AVTZ (Eh)        -75.645398
RCCSD(T)/AVQZ-sp (Eh)     -75.664289
RCCSD(T)/AV5Z-sp (Eh)     -75.670403
RCCSD(T)/CBS-sp (Eh)      -75.673329
ZPE (kcal/mol)              5.360
Cartesian coordinates in Angstroms:
 O     0.000000     0.000000    -0.057680
 H     0.000000     0.000000     0.915578
--------------------------------------------
```

**HCl($^1\Sigma^+$)** [HCl_1Sigma+.xyz]

```
--------------------------------------------
RCCSD(T)/AVTZ (Eh)       -460.345710
ZPE (kcal/mol)              4.360
Cartesian coordinates in Angstroms:
 H     0.000000     0.000000    -1.240113
CL     0.000000     0.000000     0.035257
```

```
--------------------------------------------
```

**$S_2$($^3\Sigma_g^-$)** [S2_3Sigma-.xyz]

```
--------------------------------------------
RCCSD(T)/AVTZ (Eh)       -795.467737
ZPE (kcal/mol)              1.030
Cartesian coordinates in Angstroms:
 S     0.000000     0.000000    -0.952889
 S     0.000000     0.000000     0.952889
--------------------------------------------
```

**HSCl($^1$A')** [HSCl_1Ap.xyz]

```
--------------------------------------------
```

```
RCCSD(T)/AVTZ (Eh)       -858.070385
RCCSD(T)/AVQZ-sp (Eh)    -858.106289
RCCSD(T)/AV5Z-sp (Eh)    -858.117638
RCCSD(T)/CBS-sp (Eh)     -858.122883
ZPE (kcal/mol)              5.940
Cartesian coordinates in Angstroms:
 H    -1.287902     0.000000    -1.215295
 S     0.043627     0.000000    -1.056748
CL    -0.002836     0.000000     0.990164
```
-----------------------------------------------

**$SCl_2(^1A_1)$** [SCl2_1A1.xyz]

-----------------------------------------------

```
RCCSD(T)/AVTZ (Eh)      -1317.205952
RCCSD(T)/AVQZ-sp (Eh)   -1317.262583
RCCSD(T)/AV5Z-sp (Eh)   -1317.280649
RCCSD(T)/CBS-sp (Eh)    -1317.289112
ZPE (kcal/mol)              1.790
Cartesian coordinates in Angstroms:
 S     0.000000    -0.871442     0.000000
CL     0.000000     0.398380     1.584806
CL     0.000000     0.398380    -1.584806
```
-----------------------------------------------

**$SSCl(^2A'')$** [SSCl_2App.xyz]

-----------------------------------------------

```
RCCSD(T)/AVTZ (Eh)      -1255.210830
ZPE (kcal/mol)              1.870
Cartesian coordinates in Angstroms:
 S     0.000000     0.418154    -1.643931
 S     0.000000    -0.769702    -0.145741
CL     0.000000     0.321421     1.636300
```

-----------------------------------------------

**$H_2O(^1A_1)$** [H2O_1A1.xyz]

-----------------------------------------------

```
RCCSD(T)/AVTZ (Eh)        -76.342332
ZPE (kcal/mol)             13.500
Cartesian coordinates in Angstroms:
 O     0.000000    -0.065940     0.000000
 H     0.000000     0.523257     0.757590
 H     0.000000     0.523257    -0.757590
```
-----------------------------------------------

**$HSOH(^1A)$** [HSOH_1A.xyz]

-----------------------------------------------

```
RCCSD(T)/AVTZ (Eh)       -474.052082
ZPE (kcal/mol)             14.290
```

```
Cartesian coordinates in Angstroms:
 H     0.017784    -1.276948     0.808265
 S     0.004429     0.046012     0.570330
 O    -0.063202    -0.011839    -1.101498
 H     0.844783     0.005173    -1.419744
-------------------------------------------
```

**ClSSCl($^1$A)** [ClSSCl_1A.xyz]

```
-------------------------------------------
RCCSD(T)/AVTZ (Eh)     -1714.966601
ZPE (kcal/mol)              2.870
Cartesian coordinates in Angstroms:
 S     0.446074     0.866195     0.738741
 S    -0.446074    -0.866195     0.738741
CL    -0.446074     2.110302    -0.695286
CL     0.446074    -2.110302    -0.695286
-------------------------------------------
```

**HSClOH-a($^2$A)** [HSClOH-1_2A.xyz]

```
-------------------------------------------
RCCSD(T)/AVTZ (Eh)      -933.756780
ZPE (kcal/mol)             15.420
Cartesian coordinates in Angstroms:
 H    -1.292967    -0.048309     0.514280
 S     0.015104    -0.285096     0.593593
 O    -0.031004     0.318709     2.195060
 H     0.857267     0.211770     2.559570
CL     0.012929     0.110173    -1.635344
-------------------------------------------
```

**HSClOH-b($^2$A)** [HSClOH-2_2A.xyz]

```
-------------------------------------------
RCCSD(T)/AVTZ (Eh)      -933.757683
ZPE (kcal/mol)             15.510
Cartesian coordinates in Angstroms:
 H    -1.267232     0.855488    -0.821607
 S     0.045189     0.580805    -0.949339
 O    -0.010491    -1.014855    -1.255985
 H    -0.059655    -1.433235    -0.374532
CL     0.001724    -0.050180     1.476949
-------------------------------------------
```

**$HSCl_2$($^2$A’)** [HSCl2_2Ap.xyz]

```
-------------------------------------------
RCCSD(T)/AVTZ (Eh)      -1317.775276
ZPE (kcal/mol)             7.10
Cartesian coordinates in Angstroms:
```

```
 S    -0.039911     0.422106     0.000000
 H     1.295128     0.408202     0.000000
CL    -0.000418    -0.198849     2.090759
CL    -0.000418    -0.198849    -2.090759
-------------------------------------------
```

**TS1** [S-Cl2_Cl-SCl_3App_ts.xyz]

```
-------------------------------------------
RCCSD(T)/AVTZ (Eh)      -1317.096366
RCCSD(T)/AVQZ-sp (Eh)  -1317.148780
RCCSD(T)/AV5Z-sp (Eh)  -1317.165241
RCCSD(T)/CBS-sp (Eh)   -1317.172778
ZPE (kcal/mol)              0.650
Cartesian coordinates in Angstroms:
CL     0.000000     0.144813    -2.105041
CL     0.000000    -0.272785    -0.024713
 S     0.000000     0.139967     2.329379
-------------------------------------------
```

**TS2** [SH+Cl2_Cl+HSCl_2A_ts.xyz]

```
-------------------------------------------
RCCSD(T)/AVTZ (Eh)      -1317.733314
RCCSD(T)/AVQZ-sp (Eh)  -1317.787722
RCCSD(T)/AV5Z-sp (Eh)  -1317.804798
RCCSD(T)/CBS-sp (Eh)   -1317.812609
ZPE (kcal/mol)              5.660
Cartesian coordinates in Angstroms:
 H    -1.251165     0.488727    -2.280951
 S     0.050867     0.162354    -2.287592
CL    -0.022502    -0.348174     0.053946
CL     0.012054     0.185648     2.103342

-------------------------------------------
```

**TS3** [H+SCl2_HSCl2_2A_ts.xyz]

```
-------------------------------------------
RCCSD(T)/AVTZ (Eh)      -1317.704238
RCCSD(T)/AVQZ-sp (Eh)  -1317.760984
RCCSD(T)/AV5Z-sp (Eh)  -1317.779090
RCCSD(T)/CBS-sp (Eh)   -1317.787574
ZPE (kcal/mol)              2.420
Cartesian coordinates in Angstroms:
 S    -0.053765    -0.859295    -0.017120
CL     0.022425     0.386227     1.609100
CL     0.004262     0.467271    -1.554583
 H     0.779641    -2.353968    -1.348463
-------------------------------------------
```

**TS4** [H+SCl2_SCl+HCl_2A_ts.xyz]

```
-------------------------------------------
RCCSD(T)/AVTZ (Eh)      -1317.699540
RCCSD(T)/AVQZ-sp (Eh)  -1317.756151
RCCSD(T)/AV5Z-sp (Eh)  -1317.774214
RCCSD(T)/CBS-sp (Eh)   -1317.782678
```

```
ZPE (kcal/mol)              2.200
Cartesian coordinates in Angstroms:
 H     0.235372     1.425204     3.121309
CL    -0.012819     0.324126     1.608777
 S     0.006660    -0.901408    -0.104513
CL    -0.000053     0.458957    -1.603179
```
-------------------------------------------

**TS5** [SCl-Cl2_Cl-SCl2_2A_ts.xyz]

-------------------------------------------

```
RCCSD(T)/AVTZ (Eh)      -1776.871970
RCCSD(T)/AVQZ-sp (Eh)  -1776.947260
RCCSD(T)/AV5Z-sp (Eh)  -1776.971115
RCCSD(T)/CBS-sp (Eh)   -1776.982179
ZPE (kcal/mol)              1.930
Cartesian coordinates in Angstroms:
CL    -0.007338    -0.793483     2.395267
 S     0.154710     0.917194     1.376143
CL    -0.276899     0.307660    -0.821060
CL     0.142786    -0.352769    -2.832417
```

-------------------------------------------

**TS6** [OH+HSCl_SCl+H2O_2A_ts.xyz]

-------------------------------------------

```
RCCSD(T)/AVTZ (Eh)       -933.716657
RCCSD(T)/AVQZ-sp (Eh)   -933.771463
RCCSD(T)/AV5Z-sp (Eh)   -933.788817
RCCSD(T)/CBS-sp (Eh)    -933.796858
ZPE (kcal/mol)             11.740
Cartesian coordinates in Angstroms:
 H     0.466376    -0.175159    -1.173830
 S    -0.011796    -0.956017    -0.121856
CL     0.003837     0.535704     1.267703
 O     0.041339     0.693469    -2.321288
 H    -0.881387     0.910320    -2.105831
```
-------------------------------------------

**TS7** [HSClOH_OH+HSCl_2A_ts.xyz]

-------------------------------------------

```
RCCSD(T)/AVTZ (Eh)       -933.720371
ZPE (kcal/mol)             12.190
Cartesian coordinates in Angstroms:
 H    -1.089020    -0.852770    -0.610364
 S     0.033544    -1.079711     0.086674
CL    -0.008219     0.710602     1.083992
```

```
 O    -0.008314     0.587075    -2.395776
 H     0.442022     1.132042    -1.728184
--------------------------------------------
```

**TS8** [HSClOH_1-2_2A_ts.xyz]

```
--------------------------------------------
RCCSD(T)/AVTZ (Eh)      -933.750650
ZPE (kcal/mol)            15.050
Cartesian coordinates in Angstroms:
 H    -1.281688    -0.638349     0.560965
 S     0.040678    -0.513006     0.753914
 O     0.002581     0.625497     1.950741
 H    -0.130696     1.488997     1.536883
CL     0.002334     0.158420    -1.642044
```